# ARE GAMMA-RAY BURSTS DUE TO ROTATION POWERED HIGH VELOCITY PULSARS IN THE HALO ?

Dieter H. Hartmann[1,3] and Ramesh Narayan[2,3]

[1] *Dept. of Physics and Astronomy, Clemson University, Clemson, SC 29634;*
*hartmann@grb.phys.clemson.edu*
[2] *Harvard-Smithsonian Center for Astrophysics, Cambridge, MA 02138;*
*narayan@cfa.harvard.edu*
[3] *Institute for Theoretical Physics, University of California, Santa Barbara, CA 93106*

## Abstract

The BATSE experiment has now observed more than 1100 gamma-ray bursts. The observed angular distribution is isotropic, while the brightness distribution of bursts shows a reduced number of faint events. These observations favor a cosmological burst origin. Alternatively, very extended Galactic halo (EGH) models have been considered. In the latter scenario, the currently favored source of gamma-ray bursts involves high-velocity pulsars ejected from the Galactic disk. To be compatible with the observed isotropy, most models invoke a sampling distance of D $\sim$ 300 kpc, a turn-on delay $t_{\rm turn-on} \sim 3 \times 10^7$ yrs, and a source life time $t_{\rm max} \sim 10^9$ yrs. We consider the global energy requirements of such models and show that the largest known resource, rotational kinetic energy, is insufficient by orders of magnitude to provide the observed burst rate. More exotic energy sources or differently tuned pulsar models may be able to get around the global energy constraint but at the cost of becoming contrived. Thus, while extended halo models are not ruled out, our argument places a severe obstacle for such models and we encourage proponents of EGH models to clearly address the issue of energetics.

# 1 INTRODUCTION

The "great debate" on the distance scale of cosmic gamma-ray bursts (GRB) (Fishman 1995; Lamb 1995; Paczyński 1995) considered two alternatives; cosmological bursts and events that occur in an extended Galactic halo. The previous paradigm of nearby Galactic neutron stars with a Population I distribution has fallen out of favor because of the combined observations of (i) an isotropic angular distribution of GRBs and (ii) reduced source counts at the faint end of the apparent flux distribution (Meegan *et al.* 1992; Briggs *et al.* 1995). The absence of even a hint of a "Milky Way" band in the GRB distribution makes it very hard to retain the hypothesis that local neutron stars provide the underlying source population. Some recent reviews of these and related issues are provided in Briggs (1995), Fishman & Meegan (1995), and Hartmann (1995).

To save the Galactic hypothesis under the tight constraint of isotropy it is necessary to consider a highly extended structure large enough to minimize the dipole due to the solar offset from the Galactic center. The current multipole limits (Briggs *et al.* 1995; Tegmark, *et al.* 1995) require galactocentric shells with typical radii $\sim 200$ kpc. On the other hand, halos that are too large will yield an excess of bursts towards M31, which is not observed (e.g., Hakkila *et al.* 1994, 1995; Briggs, *et al.* 1995). Because of these twin constraints, most halo models invoke a limiting sampling distance of about 300 kpc for the BATSE bursts.

Any source population that couples dynamically to the Galactic potential is likely to have a radial profile similar to the Hubble profile frequently employed in descriptions of the Galactic dark matter halo. The studies by Hakkila *et al.* (1994, 1995) rule out such a source density distribution for GRBs since it would lead to a detectable anisotropy. Similarly, scenarios that involve old sources trapped in the Galactic potential would also show strong anisotropies, at least in the inner regions of the Galaxy. An example of the latter situation is a population of neutron stars born in the halo (Eichler & Silk 1992; Hartmann 1992), which generates a halo distribution that compares poorly with BATSE data (Briggs *et al.* 1995). High velocity objects (neutron stars) that left the Galaxy a long time ago are distributed roughly uniformly between galaxies and would constitute a cosmological rather than a Galactic model, although the distance scale would be intermediate between the halo scale (300 kpc) and the cosmological scale (3,000 Mpc) (Hartmann

1995).

Currently, the only surviving Galactic halo model is one in which the bursts are produced by high velocity sources born in the vicinity of the disk streaming out into the halo (Li & Dermer 1992; Bulik & Lamb 1995; Podsiadlowski *et al.* 1995). We consider this scenario in this paper and for definiteness assume that the relevant sources are high velocity pulsars (HVPs). The recent upward revision of radio pulsar velocities (Lyne & Lorimer 1994) apparently provides some support and motivation for such a scenario.

Some general arguments can be made about the requirements of a HVP model. First, one demands of course that essentially no pulsars with small velocities produce observable GRBs, else their strong concentration to the Galactic disk would quickly violate the isotropy constraints. HVP models come in two flavors. In "free-streaming" models, the pulsars receive at formation velocity kicks in excess of the escape velocity of the Galaxy and stream away on more-or-less radial orbits. In such models, the minimum velocity kick of the pulsars is $\sim 800$ km s$^{-1}$. This is a little larger than the typical escape velocity of $v_{\rm esc} \sim 600$ km s$^{-1}$ in order to compensate for cases where part of the kick is cancelled by orbital motion in the Galactic disk. In the alternate "quasi-virialized" halo models, some fraction of the stars remain bound to the Galaxy, and so the sources have somewhat smaller velocities. However, since the trapped stars need to fill a volume out to 300 kpc in the halo their velocities cannot be much less than escape. In our discussions, we assume a typical velocity $v \sim 10^3 v_3$ km s$^{-1}$.

All HVP models require a delayed turn-on of observable GRB activity in the sources. The delay could be due to either a true physical delay in the burst mechanism within the source, or a correlation between beaming and direction of kick such that only bursts which have traveled a large distance from their point of origin are visible to us (Li, Duncan, & Thompson 1994; Li & Duncan 1995). In either case, the bursts visible to us can only originate from galactocentric radii greater than $\sim 30$ kpc, since otherwise there would be an overproduction of bursts near the disk and a violation of isotropy. To get out to a distance of $\sim 30$ kpc requires a turn-on delay of $t_{\rm turn-on} \sim 3 \times 10^7/v_3$ yrs. In the free streaming scenario, these pulsars are visible for a total duration $t_{\rm max} \sim 3 \times 10^8/v_3$ yrs, after which they travel beyond our sampling distance of 300 kpc. In the quasi-virialized scenario, the pulsars may be visible for a slightly longer time, but not longer than $t_{\rm max} \sim 10^9$ yrs since otherwise they would virialize too well and violate the isotropy constraint.

The outline of the paper is as follows. We use the BATSE observations (Meegan *et al.* 1995b,c) along with the above constraints to show that the energy emitted per burst in the HVP model is $\sim 10^{43}$ ergs. We then combine this with the known birthrate of radio pulsars to estimate that each high velocity pulsar in the Galaxy must emit $\sim 10^{49}$ ergs in the form of GRBs during its lifetime. This is an enormous amount of energy and we show that in most scenarios the available rotational kinetic energy is insufficient by orders of magnitude. We therefore conclude that, within the halo model, GRBs must be powered by an energy source other than rotation. We discuss some possibilities.

## 2 GLOBAL ENERGY CONSTRAINTS

The very first check on any GRB model is verification of the energy scale: can a mechanism generate enough impulsive energy to explain the observed fluxes? The next step usually involves the question of isotropy and brightness distributions. Another check, at least as important as geometric arguments, involves the global energy required to sustain burst activity. While cosmological models usually invoke singular events, such as the merger of two compact objects (e.g., Narayan, Paczyński & Piran 1992; Meszaros & Rees 1993), Galactic models require multiple outbursts from each source. The energy required for these repeat events can provide severe constraints on burst models as was demonstrated in the framework of the old nearby neutron star paradigm by Blaes *et al.* (1989, 1990) and Madau (1992). We repeat this exercise for the HVP scenario of Galactic GRBs.

We consider the following model. Some fraction of pulsars are born with high velocities characterized by the parameter $v_3$. We write this fraction as $0.1 f_v v_3^{-\alpha}$ in the reasonable expectation that roughly 10% of radio pulsars are formed with velocities in excess of $10^3$ km s$^{-1}$ and that the fraction decreases with increasing velocity as some power of $v_3$. The actual fraction of HVPs with velocity $v_3$ is poorly known, because we are dealing with the tail of the distribution function (e.g., Lyne & Lorimer 1994) that currently contains only a few sources. While some authors would argue for $f_v$ to be as high as 5 (Lamb 1995), others may feel that this parameter is significantly less than unity (Hartmann & Greiner 1995). We feel a fiducial value is $f_v = 1$.

Currently, there is no observational estimate of $\alpha$, except that it must clearly be a positive number.

Next, we assume that some fraction $f_\gamma$ of the HVPs actually become GRB sources. In the standard HVP scenario, $f_\gamma = 1$. However, velocity may not be the only selection criterion for GRB activity, as we will argue below, and $f_\gamma$ could be significantly less than unity.

Radio pulsars are born in the Galaxy at the rate of roughly one pulsar every 100 years (Lyne, Manchester & Taylor 1985; Narayan & Ostriker 1990). Let us write this rate as $R_b = 10^{-2} r_{-2}$ yr$^{-1}$. The rate at which GRB-producing neutron stars are born is then $10^{-3} r_{-2} f_v f_\gamma v_3^{-\alpha}$ yr$^{-1}$. On the other hand, BATSE has been detecting GRBs at the rate of approximately one per day. After correcting for Earth blockage and temporal gaps, the inferred all-sky GRB rate at the BATSE sensitivity level is $\sim 10^3$ yr$^{-1}$. Combining these two rates, we find that in steady state, each bursting HVP must produce

$$N_{\text{bursts}} \sim 10^6 \ (r_{-2} \ f_v \ f_\gamma)^{-1} \ v_3^\alpha \qquad (1)$$

GRBs during its lifetime. The steady state assumption on which equation (1) is based is quite reasonable since we argued in § 1 that the oldest bursting sources cannot be much older than $10^9$ yrs. It is not expected that the Galactic rate of formation of neutron stars has evolved significantly over such a short period. If we allow ourselves to make use of all neutron stars ever formed in the Galaxy, then the time-averaged effective birthrate could be increased by one order of magnitude, but not much more (cf. Timmes, Woosley & Weaver 1995, 1996), and the number of burst repetitions per source could be reduced by an equivalent factor. In any case, it is clear that the HVP scenario predicts a large number of burst repetitions per source.

The faintest bursts detected by BATSE have a peak intensity of $10^{-7}$ ergs cm$^{-2}$ s$^{-1}$. Using a mean effective burst duration (derived from the ratio of fluence to peak flux) of $T_{\text{eff}} \sim 10$ s (Lee & Petrosian 1995), the average fluence from the most distant bursts detected is $\sim 10^{-6}$ ergs cm$^{-2}$. Let us assume that the distance to the dimmest sources is $D = 300$ $D_{300}$ kpc. The acceptable range of $D_{300}$ is $\sim 0.5 - 2$, since outside of this range anisotropies quickly exceed the observational limits. With these estimates, the mean energy emitted in each GRB is

$$E_0 \sim 10^{43} D_{300}^2 \ \text{ergs} \ . \qquad (2)$$

Combining this with the burst repetition estimate in (1), we find that the total energy emitted in the form of observable GRBs by each source is

$$E_{\rm tot} \sim 10^{49} \; {\rm D}_{300}^2 \; (r_{-2} \; f_{\rm v} \; f_\gamma)^{-1} \; v_3^\alpha \; {\rm ergs} \; . \qquad (3)$$

With reasonable (perhaps optimistic) choices of the parameters, we thus find as a benchmark that we need to produce $\sim 10^{49}$ ergs per source. We discuss in succeeding subsections if this much energy is likely to be available from HVPs. Note that if the product $f_{\rm v} f_\gamma$ is much less than unity, which it very well might be, the energy requirement is even more extreme.

While the energy argument is the primary thrust of this paper, we mention here another interesting issue related to the repetition rate of the sources. In the free-streaming model, the sources are visible within the BATSE sampling distance for only a period of $t_{\rm max} \sim 3 \times 10^8 {\rm D}_{300} v_3^{-1}$ yrs, while in the quasi-virialized model the period is $t_{\rm max} \sim 10^9$ yrs. We therefore expect the average source to repeat at the following rates in the two scenarios:

$$r = \frac{N_{\rm bursts}}{t_{\rm max}} \sim \; 3 \times 10^{-3} \; {\rm D}_{300}^{-1} \; (r_{-2} \; f_{\rm v} \; f_\gamma)^{-1} \; v_3^{\alpha+1} \; {\rm yr}^{-1} \qquad {\rm (streaming)} \quad (4)$$

$$r \sim 10^{-3} \; (r_{-2} \; f_{\rm v} \; f_\gamma)^{-1} \; v_3^\alpha \; {\rm yr}^{-1} \; . \qquad {\rm (quasi-virialized)} \qquad (5)$$

The current data limit the fraction of sources that repeat within 5 years to less than a few percent (Hartmann *et al.* 1995), somewhat smaller than the limit of $\sim 10-20\%$ derived for 2B data (e.g, Meegan *et al.* 1995a). The limits are improving steadily and could lead eventually to a serious inconsistency if some of the uncertain factors like $f_{\rm v}$ or $f_\gamma$ are made too small.

## 3   ROTATION-POWERED PULSARS

It is believed that radio radiation as well as higher energy (optical to $\gamma$-ray) radiation from pulsars is ultimately powered by the rotational kinetic energy of the neutron star. It is therefore natural to assume that the same energy source also powers $\gamma$-ray bursts in the halo HVP model. At the simplest level, if we assume that the initial rotation period of pulsars is typically $P(0) \sim 0.01$ s (using the Crab pulsar as the prototype), the initial rotational kinetic energy is $\sim 2 \times 10^{50}$ ergs (for a moment of inertia I $= 10^{45}$ g cm$^2$), which would appear to be comfortably larger than the energy of $10^{49}$ ergs required to power the GRBs. This argument is however misleading and we show now that there is in fact a very serious problem with using rotational energy to power GRBs.

Let us write the surface dipole magnetic field of the pulsar as $B = 10^{12} B_{12}$ G and express time in units of $10^8$ yrs, $t = 10^8 \, t_8$ yrs. If electromagnetic dipole radiation dominates the spin-down torque of the neutron star the period derivative (in units of $10^{-15}$ s s$^{-1}$) is given by the standard relation

$$\dot{P}_{-15} = P^{-1} \, B_{12}^2 \; . \tag{6}$$

If we ignore field decay, this equation integrates to give

$$[P(t_8)]^2 = [P(0)]^2 + 6 \, B_{12}^2 \, t_8 \; . \tag{7}$$

Let us be conservative and set $P(0) = 0$. As we discussed in § 1, the escaping pulsars in the HVP scenario turn on as GRB sources only after they travel a distance $\sim 30$ kpc from their point of origin. The time to travel this distance is $t_{\rm turn-on} \sim 3 \times 10^7 v_3^{-1}$ yrs. If we substitute this time in equation (7) and assume (i) that a fraction $\xi$ of the rotational energy is converted into $\gamma$-rays, and (ii) make the conservative assumption that all the bursts occur while the pulsar is within the BATSE sampling distance of 300 kpc, then the maximum energy in observable GRBs that we can expect from the pulsar is

$$E_{\rm max} = \frac{\xi}{2} I \left[ \frac{2\pi}{P(t_{\rm turn-on})} \right]^2 \sim 10^{46} \, \xi B_{12}^{-2} v_3 \; \text{ergs} \; . \tag{8}$$

Even with an optimistic choice of $\xi = 1$, we see that for a typical radio pulsar with $B_{12} > 1$ the available rotational energy is less than the required $10^{49}$ ergs by orders of magnitude.

Are there any pulsars which can provide $10^{49}$ ergs of rotational energy ? Figure 1 shows the two-dimensional plane of pulsar field strength $B_{12}$ and turn-on time $t_{\rm turn-on}$ and displays contours of available rotational energy $E_{\rm max}$, assuming $\xi = 1$. Various studies of radio pulsars (e.g., Narayan & Ostriker 1990; Bhattacharya et al. 1992) have shown that the typical distribution of magnetic field strengths of radio pulsars at birth is fairly narrow (in the log). The peak of the distribution is around $\log(B) \sim 12.3 - 12.4$ and the dispersion around the mean is $\sigma_{\log B} \sim 0.3 - 0.4$. Figure 1 shows that pulsars with such field strengths lose their rotational energy quite rapidly and need to turn on as GRBs in less than $10^5$ yrs, otherwise they just do not have enough energy left. Even at a velocity of $10^3$ km s$^{-1}$, pulsars travel only 100 pc in this time and this is inconsistent with the observed isotropy.

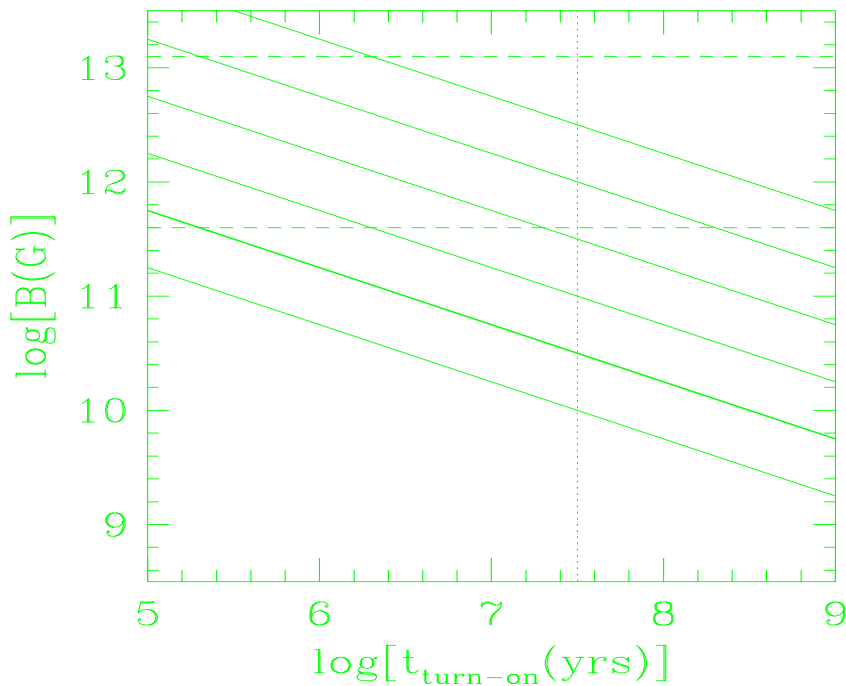

Figure 1: The solid lines show, on the two-dimensional plane of turn-on time $t_{\text{turn-on}}$ and pulsar magnetic field $B$, contours indicating the available rotational kinetic energy. From below, the lines correspond to $10^{50}$, $10^{49}$, $10^{48}$, $10^{47}$, $10^{46}$, $10^{45}$ ergs, respectively. The energy required to power gamma-ray bursts in a canonical halo model is $\sim 10^{49}$ ergs (§ 2) and is shown by a thicker line. The two horizontal dashed lines are a somewhat generous representation of the range of magnetic field strengths of radio pulsars at birth ($\pm 2\sigma$ from Narayan & Ostriker 1990 and Bhattacharya et al. 1992). The vertical dotted line is the canonical delayed turn-on time in HVP models. The intrinsic dilemma of halo models is clearly seen. Pulsars with canonical field strengths $> 10^{12}$ G have sufficient rotational energy to power the observed GRBs only if they turn on as bursters less than $10^5$ yrs after they are born, but this is in conflict with the observed isotropy of the sources. If the turn-on time is taken to be $10^{7.5}$ yrs in order to fit the isotropy, then only pulsars with fields less than $10^{10.5}$ G will have enough energy, but there are very few pulsars with such field strengths.

On the other hand, if we take $t_{\rm turn-on} \sim 3 \times 10^7$ yrs, as proposed in most HVP models, then we see from Fig. 1 that the field strength has to be lower than $10^{10.5}$ G if the pulsar is to retain enough energy by the time it turns on. Figure 2 shows all known pulsars, except those in globular clusters, plotted in the $BP$ plane. (We used the updated version of the catalog published by Taylor, Manchester & Lyne 1993, which is current as of May 3, 1995.) If we leave aside binary and millisecond pulsars, we see that there is not a single regular pulsar which has a field strength below $10^{10.5}$ G and which has a rotational kinetic energy of $10^{49}$ ergs. Therefore, the birthrate of such systems must be extremely small. This means that the parameter $f_\gamma$ which we introduced in the previous subsection is much less than unity, which then implies that the required energy per source is not $10^{49}$ ergs, but very much larger. Going back to Fig. 1, this means that the field strength must be much smaller than $10^{10.5}$ G which of course reduces the birthrate still further. By carrying through this argument it is easy to convince oneself that, given what we know about regular radio pulsars, it is impossible to reconcile the energy requirement of equation (3) with the limit on the available energy given in equation (8).

Millisecond and binary pulsars do have the kind of rotation rates and field strengths required by equation (8). However, the birthrates of these objects have been estimated to be orders of magnitude less than the $10^{-2}$ yr$^{-1}$ assumed in § 2 (Kulkarni & Narayan 1988, Narayan, Piran & Shemi 1991). This means that the energy per source has correspondingly to be orders of magnitude greater than $10^{49}$ ergs and once again we run into a spiraling inconsistency. Further, there is no evidence that binary and millisecond pulsars have velocities $\sim 10^3$ km s$^{-1}$ as required in the HPV model. Indeed, van Paradijs & White (1995) find that low-mass X-ray binaries, the supposed progenitors of binary and millisecond pulsars, have a scale height of only 0.5 kpc. It is impossible for such a parent population to produce a population of HVPs extending out to 300 kpc, as required in the halo GRB model.

The above analysis was based on the assumption that fields do not decay. If instead we assume that the field strength decays exponentially with time

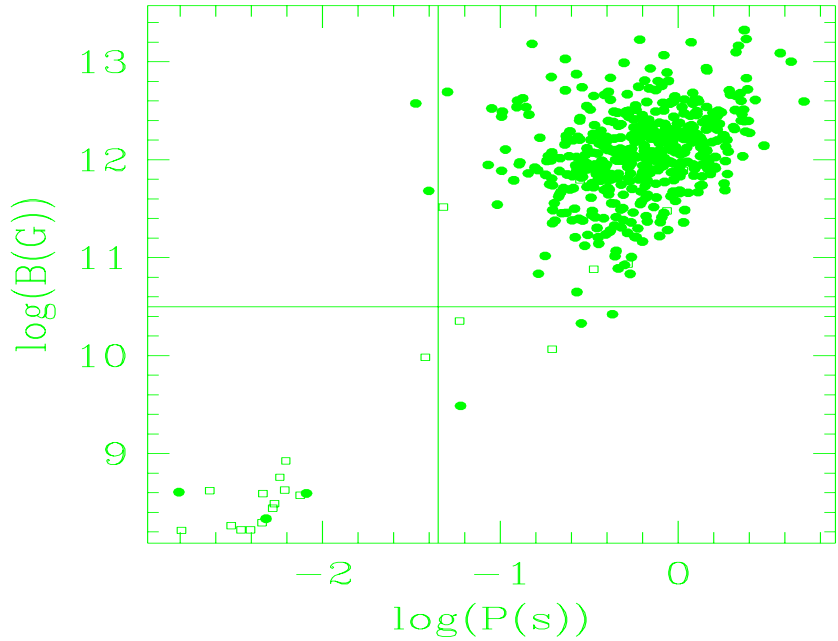

Figure 2: The solid circles and open squares show known single and binary radio pulsars, plotted on the $BP$ diagram. Pulsars in globular clusters have not been shown as they are not relevant to the HVP model of GRBs. The vertical line corresponds to a rotational kinetic energy of $10^{49}$ ergs, and the horizontal line corresponds to a field strength of $10^{10.5}$ G. In the HVP model of GRBs, only pulsars below and to the left of these lines are suitable progenitors of burst sources. Note that this region is dominated by binary pulsars and three single millisecond pulsars. These sources have a low birthrate and low velocities and, for these reasons, are ruled out as sources of GRBs.

with a time constant $\tau = 10^8 \tau_8$ yrs, then equation (6) is modified to

$$\dot{P}_{-15} = P^{-1} \, B_{12}^2 \exp(-2t_8/\tau_8) \,, \tag{9}$$

which gives (setting $P(0) = 0$, as before)

$$[P(t_8)]^2 = 3 \, B_{12}^2 \, \tau_8 [1 - \exp(-2t_8/\tau_8)] \,. \tag{10}$$

At first sight this appears to help since the rotation period at a given time $t_8$ is shorter in the presence of field decay than without it, so that the pulsar can have a larger kinetic energy when it turns on. However, when there is exponential field decay, the pulsar does not spin down all the way to zero rotation, but asymptotically approaches a limiting period of $B_{12}(3\tau_8)^{1/2}$ as $t \to \infty$ (eq. 10). Therefore, the maximum rotational energy which is available via spin-down at a given turn-on time is given by

$$E_{\max} = \frac{2\pi^2 \xi I}{3 B_{12}^2 \tau_8} \left[ \frac{1}{1 - \exp(-2t_8/\tau_8)} - 1 \right] \,. \tag{11}$$

This estimate of $E_{\max}$ is lower than the no-decay estimate given in equation (8) for all choices of $t_8$ and $\tau_8$. Thus, having no field decay is the most optimistic assumption we can make in the HVP picture.

There is only one scenario that we can think of in which rotational energy could provide the energy for halo GRBs. Let us suppose that there is a large population of neutron stars which are spun-up by accretion in a binary, and whose magnetic field strengths, though initially large, are decreased in the process of accretion (cf. Shibazaki *et al.* 1989; Romani 1990). In this scenario, the spun-up pulsars will be reborn as low field, fast-spinning objects, precisely what is needed for the HPV model. The main difficulty with this scenario is that we have no evidence for a population of high velocity single pulsars with the properties of these spun-up pulsars (see Fig. 2). As already mentioned, many of the binary pulsars and all millisecond pulsars do in fact correspond to such objects, but there are not enough of them to satisfy the energy constraint and they do not have the right velocities.

## 4  OTHER ENERGY SOURCES

Are there good alternative energy sources for GRBs from HVPs in the halo? If accretion of external matter is the source (Colgate & Leonard 1994, 1995;

Woosley 1993, 1994) approximately $10^{-4}$ $M_\odot$ of matter must be stored around the neutron star and accreted in order to release $\sim 10^{49}$ ergs at an assumed efficiency of 10%. The accretion will have to occur in $\sim 10^6$ discrete installments of $\sim 10^{23}$ g each. Thus each accreting blob must be significantly larger than a typical large comet (e.g. Halley's comet) in the solar system; the chunks of material resemble more closely planetesimals.

The formation of an accretion disk around a HVP moving through the ejecta of the supernova event was suggested by Lin, Woosley, & Bodenheimer (1991), who also showed that planetesimals with the right properties could be formed. Woosley & Herant (1995) support this accretion picture with detailed SPH calculations. One added advantage of this model is that the accretion rate, and thus presumably the GRB rate, anti-correlates with the spin-down of the pulsar. As the pulsar slows down planetesimals can accrete with higher efficiency. This provides a physical argument in support of a delayed turn-on. On the other hand, Woosley (1993) estimates that burst activity would continue (perhaps even at a higher rate) for $10^{10}$ yrs. Since the observable bursts cut-off after about $10^9$ yrs for the reasons mentioned earlier, this increases the energy requirements significantly. Accretion disk evolution would spin down the pulsar faster than predicted by dipole radiation (eq. 6) and thus reduce the available rotational energy (Woosley & Herant 1995). If an accretion disk provides the energy for GRB activity the role of rotational energy is diminsihed.

While an HVP may be able to form a disk that contains massive enough planetesimals it may fall short in the required total mass by an order of magnitude (Woosley & Herant 1995). Furthermore, the tidal disruption of the planetoids would typically occur on timescales much longer than typical burst durations, depending on the distribution of orbital parameters. In addition, Woosley & Herant (1995) estimate the energy conversion efficiency to be $\sim 10^{-3}$, two orders of magnitude less than our canonical value. It seems that even accretion disks around HVPs face a severe energy budget deficit.

Direct accretion from the ISM in the halo is not efficient because of the low density of the medium and the high velocities of the stars, although the accretion rate can in fact be enhanced above the Bondi-Hoyle hydrodynamic rate (Harding & Leventhal 1992).

Phase transitions in neutron star interiors (e.g., Woosley 1995) may provide enough energy for a few bursts, but do not seem capable of producing the required rate per source. Elastic energy stored in the neutron star crust is estimated to be less than $10^{44}$ ergs (Blaes *et al.* 1989, 1990; Ruderman 1991; Madau 1992), and thus falls far short of the required amount.

Magnetic energy is only sufficient if each neutron star contains an internal field approaching $10^{16}$ Gauss. Even if we ignore the complicated question of the mechanisms of releasing this energy intermittently at a steady rate over $10^{8-9}$ yrs, there is no compelling reason to believe that all pulsars have such internal fields.

External fields of pulsars are typically $\sim 10^{12} - 10^{13}$ G and are much too small to account for an energy budget of $10^{49}$ ergs. It has been suggested that a correlation between field strength and pulsar velocity (Anderson & Lyne 1983; Stollman & van den Heuvel 1986; Bailes 1989) may provide the explanation for why only HVPs become GRBs (Li & Dermer 1992). However, recent studies by Lorimer, Lyne, & Anderson (1995) suggest only marginal evidence for such a correlation in the pulsar data. In any case, if we require ultra-high external fields for GRB activity, then we know that the fraction of pulsars endowed with those fields is very small. This means that $f_\gamma \ll 1$ which reduces the number of available sources and increases the burst repetition rate per source. The required energy would be correspondingly larger and this would demand higher fields, etc. As before, the resulting runaway inconsistency argues against producing GRBs from a small population of magnetars in the halo (Duncan & Thompson 1992). Furthermore, the increasing specific burst rate per source would eventually violate the recurrence limits of BATSE (e.g., Meegan *et al.* 1995a).

# 5 DISCUSSION

The global energy budget requirements for burst models that invoke high velocity pulsars are so demanding that this "simple" scenario encounters a serious shortage of energy. Our estimates suggest that GRBs that draw their energy from the spin down of the pulsar do not have enough resources to provide the large number of events per source required to explain the observed burst rate. Even with a spin period as short as $P = 0.1$ s, the total rotational energy of a pulsar is just barely compatible with the energy required to power the observed GRBs. Since the halo HVP model requires a substantial delay in the turn-on of GRB activity on the order of $t_{\text{turn-on}} \sim 3 \times 10^7$ yrs, the actual rotation period at the time of turn-on is much slower than 0.1 s, and consequently the available rotational energy falls short by many orders of magnitude (see Fig. 1). The question that naturally arises is: should we completely abandon Galactic halo models or are there reasonable ways out of the global energy crisis ?

The fundamental energy problem emphasized in this work arises because of the small birth rate of sources ($< 10^{-2}$ per year) compared to the high rate of GRBs ($\sim 10^3$ yr$^{-1}$). Circumnavigating the energy crisis requires either more sources or much lower energies per event. If we continue relying on neutron stars ("dead pulsars") as the underlying population, there is very little freedom regarding source numbers. Some authors have suggested the possibility that neutron stars born during the formation phase of the Galaxy could number $\sim 10^{10}$ (Gurevich, *et al.* 1993), but constraints from chemical evolution arguments argue against such an abundant population of relic neutron stars (Hartmann 1992; Eichler & Silk 1992; Timmes *et al.* 1995, 1996). Hartmann *et al.* (1994) estimate that perhaps $10^5$ to $10^6$ neutron stars born in the halo are available as GRB sources, which is comparable to the number found for the HVP scenario. Furthermore, the spatial distribution of such relic neutron stars would resemble that of the dark matter halo, which is inconsistent with the observed isotropy.

Failing to provide more sources leads to consideration of beamed emission as a way out of the energy crisis. Random geometric beaming does not help because the decrease in energy per event is exactly compensated by the increase in the specific burst rate needed to generate the observed burst frequency. The way around this dilemma is some form of special beaming. In the model of Li, Duncan, & Thompson (1994) the emission is beamed along the direction of motion of the pulsar. Although the physical reasons for this effect are not well established (but see Duncan, Li, & Thompson 1993; Li, Duncan, & Thompson 1994) we consider the implications of such a correlation. First of all, this beaming pattern reduces the detection probability of sources close to the disk. This removes the undesirable anisotropy of young HVPs without the need for a delayed turn-on. The further the star moves from the Galactic center region the more likely its detection becomes. Eventually we see every burst from these HVPs. While this argument reduces the isotropy constraints, it has very little effect on the required burst rate estimate because the observed rate is dominated by the large number of distant sources, all of which we can detect. However, it does reduce the amount of energy per event by the beaming fraction $f_b$. If bursts become visible when the sources have traveled $\sim 30$ kpc, we estimate $f_b \sim 0.1$ so that the total energy needed is $10^{48}$ ergs rather than $10^{49}$ ergs. As equation (8) shows, this is still much too large to be provided by spin-down.

One might instead try to adjust the beaming factor to be $f_b \sim 10^{-3}$ so as to bridge the gap between equations (3) and (8) with a standard pulsar magnetic field strength of $B_{12} \sim 1$. This implies a beaming cone with an angle $\sim 1-2$ degrees, which means that most sources do not turn on as observable GRB sources at Earth until they travel 300 kpc from the disk. This of course means that $t_{\text{turn-on}}$ increases to $\sim 3 \times 10^8/v_3$ yrs. The rotational energy $E_{\max}$ then decreases by another factor of 10 (eq. 8) and we are left still with a serious discrepancy. It appears that there is no way to arrange beaming so as to produce GRBs from HVPs with typical $B_{12} \sim 1$ fields. As we have shown, including field decay only makes matters worse.

Within the rotation-powered HVP scenario, there is only one model we could come up with which could satisfy the global energy constraint. We could imagine that a substantial fraction of the neutron stars in the Galaxy are spun-up to short rotation periods via binary evolution and then ejected into the halo with high velocities. If we further postulate that these neutron stars have their magnetic fields decreased below $10^{10.5}$ G as a result of accretion and if we assume that once the neutron star is ejected the field does not decay any further, then we would have a population with exactly the properties needed (according to Fig. 1) to explain the observed bursts. The only pulsars known with these properties are binary pulsars, and estimates of the birthrates of these systems suggest that they represent only a small fraction of the neutron stars born in the Galaxy. There is no evidence for an invisible population of such stars with the kind of birthrate $\sim 10^{-2}$ yr$^{-1}$ needed to explain halo GRBs. Nevertheless, one could in principle arbitrarily assume that such systems exist and that they do not become radio pulsars so that they are more numerous than we think.

There is one interesting problem associated with these radio-silent neutron stars, namely that the oriented beaming scenario of Li, Duncan & Thompson (1994) cannot be applied to them. Recall that in this scenario, one assumes that the beam is aligned in some way with the direction of kick imparted to the neutron star by an asymmetric supernova explosion. Now, the neutron stars in question remain in binaries after they are formed and therefore largely lose memory of their direction of kick as they continue orbiting around their companions. When one of these stars is finally ejected, it is as a result of the explosion of the companion star, and there will be very little correlation between the final direction of motion of the older spun-up neutron star and its original kick velocity. Therefore, one does not expect the beam of the older neutron star (which is the one we wish to use to produce GRBs) and its direction of motion. If these stars begin bursting soon after they are formed then they will be visible while they are still close to the disk and it will be impossible to explain the isotropy of the bursts. Only with a true physical delayed turn-on could one explain the isotropy and this leaves us with the problem of explaining the reason for the delay.

Our conclusion is that it is essentially impossible to explain GRBs as rotation-powered HVPs in the Galactic halo. If we wish to retain the halo hypothesis then we must consider some of the alternate sources of energy discussed in § 4. Among the ideas discussed there, accretion-power via orbiting planetisimals (Epstein 1985; Colgate & Leonard 1994, 1995; Woosley 1994; Lin, Woosley, & Bodenheimer 1991; Woosley & Herant 1995) is perhaps more promising than the others. This model has not been worked out in enough detail and it remains to be seen if a large enough fraction of neutron stars can retain sufficiently extensive clouds of planetesimals as they are ejected into the halo. It is our suspicion that this model too will become highly contrived by the time it is put to the combined tests of isotropy, source counts and the global energy constraint we have highlighted in this paper.

The authors thank all participants of the ITP workshop on Gamma-Ray and Other Non-Thermal Sources in Astrophysics for stimulating discussions on this subject. This work was partially supported by NASA grant NAG 5-1578 at Clemson University, NSF grant AST 9423209 at the Center for Astrophysics, and NSF grant PHY94-07194 at the Institute for Theoretical Physics at UC Santa Barbara.